\documentclass[8.5pt,twoside,twocolumn]{article}
\usepackage[super,sort&compress,comma]{natbib}
\usepackage{mhchem}
\usepackage{times,mathptmx}
\usepackage{amsmath}
\usepackage{sectsty}
\usepackage{balance}

\usepackage{graphicx} %eps figures can be used instead
\usepackage{lastpage}
\usepackage[format=plain,justification=raggedright,singlelinecheck=false,font=small,labelfont=bf,labelsep=space]{caption}
\usepackage{fancyhdr}
\begin{document}

\thispagestyle{plain}
\fancypagestyle{plain}{
%\fancyhead[L]{\includegraphics[height=8pt]{headers/LH}}
%\fancyhead[C]{\hspace{-1cm}\includegraphics[height=20pt]{headers/CH}}
%\fancyhead[R]{\includegraphics[height=10pt]{headers/RH}\vspace{-0.2cm}}
\renewcommand{\headrulewidth}{1pt}}
\renewcommand{\thefootnote}{\fnsymbol{footnote}}
\renewcommand\footnoterule{\vspace*{1pt}%
\hrule width 3.4in height 0.4pt \vspace*{5pt}}
\setcounter{secnumdepth}{5}

\makeatletter
\def\subsubsection{\@startsection{subsubsection}{3}{10pt}{-1.25ex plus -1ex minus -.1ex}{0ex plus 0ex}{\normalsize\bf}}
\def\paragraph{\@startsection{paragraph}{4}{10pt}{-1.25ex plus -1ex minus -.1ex}{0ex plus 0ex}{\normalsize\textit}}
\renewcommand\@biblabel[1]{#1}
\renewcommand\@makefntext[1]%
{\noindent\makebox[0pt][r]{\@thefnmark\,}#1}
\makeatother
\renewcommand{\figurename}{\small{Fig.}~}
\sectionfont{\large}
\subsectionfont{\normalsize}

\fancyfoot{}
%\fancyfoot[LO,RE]{\vspace{-7pt}\includegraphics[height=9pt]{headers/LF}}
%\fancyfoot[CO]{\vspace{-7.2pt}\hspace{12.2cm}\includegraphics{headers/RF}}
%\fancyfoot[CE]{\vspace{-7.5pt}\hspace{-13.5cm}\includegraphics{headers/RF}}
\fancyfoot[RO]{\footnotesize{\sffamily{1--\pageref{LastPage} ~\textbar  \hspace{2pt}\thepage}}}
%\fancyfoot[LE]{\footnotesize{\sffamily{\thepage~\textbar\hspace{3.45cm} 1--\pageref{LastPage}}}}
\fancyhead{}
\renewcommand{\headrulewidth}{1pt}
\renewcommand{\footrulewidth}{1pt}
\setlength{\arrayrulewidth}{1pt}
\setlength{\columnsep}{6.5mm}
\setlength\bibsep{1pt}

\twocolumn[
  \begin{@twocolumnfalse}
\noindent\LARGE{\textbf{Why not use the thermal radiation for nanothermometry?}}
\vspace{0.6cm}

\noindent\large{\textbf{Liselotte Jauffred$^a$}}\vspace{0.5cm}
%Please note that \ast indicates the corresponding author(s) but no footnote text is required.

%\noindent\textit{\small{\textbf{Received Xth XXXXXXXXXX 20XX, Accepted Xth XXXXXXXXX 20XX\newline
%First published on the web Xth XXXXXXXXXX 200X}}}

%\noindent \textbf{\small{DOI: 10.1039/b000000x}}
\vspace{0.6cm}
%Please do not change this text.

\noindent 
\normalsize{
The measurement of temperature with nanoscale spatial resolution is an emerging new technology and it has important impact in various fields. An ideal nanothermometer should not only be accurate, but also applicable over a wide temperature range and under diverse conditions. Furthermore, the measurement time should be short enough to follow the evolution of the system. However, many of the existing techniques are limited by drawbacks such as low sensitivity and fluctuations of fluorescence. Therefore, Plank's law offers an appealing relation between the absolute temperature of the system under interrogation and the thermal spectrum. Despite this, 
thermal radiation spectroscopy is unsuitable for far-field nanothermometry, primarily because of the power loss in the near surroundings and a poor spatial resolution. 
}
\vspace{0.5cm}
 \end{@twocolumnfalse}
  ]

\footnotetext{\textit{$^{a}$The Niels Bohr Institute, University of Copenhagen, Blegdamsvej 17, Denmark.; E-mail: jauffred@nbi.dk.}}

\noindent
In 2015,  stable aerosol trapping of individual metallic nanoparticles (80-200 nm) under atmospheric pressure was reported \cite{Jauffred2015}. As the thermal conductance of air is much lower than the conductance of water, the heating associated with laser irradiation of airborne metallic nanoparticles is expected to be tunable in the range from room temperature to the melting point of gold (1,337 K). Currently, there exists no method to measure the temperature of aerosols of gold nanoparticles. Therefore, I looked into the possibility of accessing the temperature through a spectral analysis of thermal radiation.

Thermal (blackbody) radiation has a spectrum that depends entirely on the temperature of the particle. The emission intensity for a specific wavelength can be calculated from Planck's law and by balancing the absorbed power with the emission power and the heat dissipation, the particle temperature can be extracted. 

To my knowledge, the first attempts to probe a temperature field at small scales were based on the use of local nanotips used as a nanoscale thermocoupler. This is the so called SThM (scanning thermal microscopy) technique and it was introduced in 2014 by the group of Levy \cite{Desiatov2014}. The authors shoved that they, with the nanotip, were able to measure temperature rises of 15 K. In 2016, the group of S{\"u}zer reported on another near-field technique to map the temperature of plasmonic nanoantennas \cite{Kinkhabwala2015}. The experiments were conducted in the context of heat-assisted magnetic recording and the technique was termed Polymer Imprint Thermal Mapping (PITM). The technique explores thermosensitive polymers that permanently cross-links upon heating, which causes a thickening that can be subsequently mapped with AFM. However, these near-field techniques are very invasive and thus has limited application \cite{Baffou2017a} particularly, for nanoparticle aerosols. In the following, I will evaluate the possibility of measuring thermal radiation of gold nanoparticles in the far-field, instead of the near-field.
\begin{center}
  \includegraphics[width=8cm]{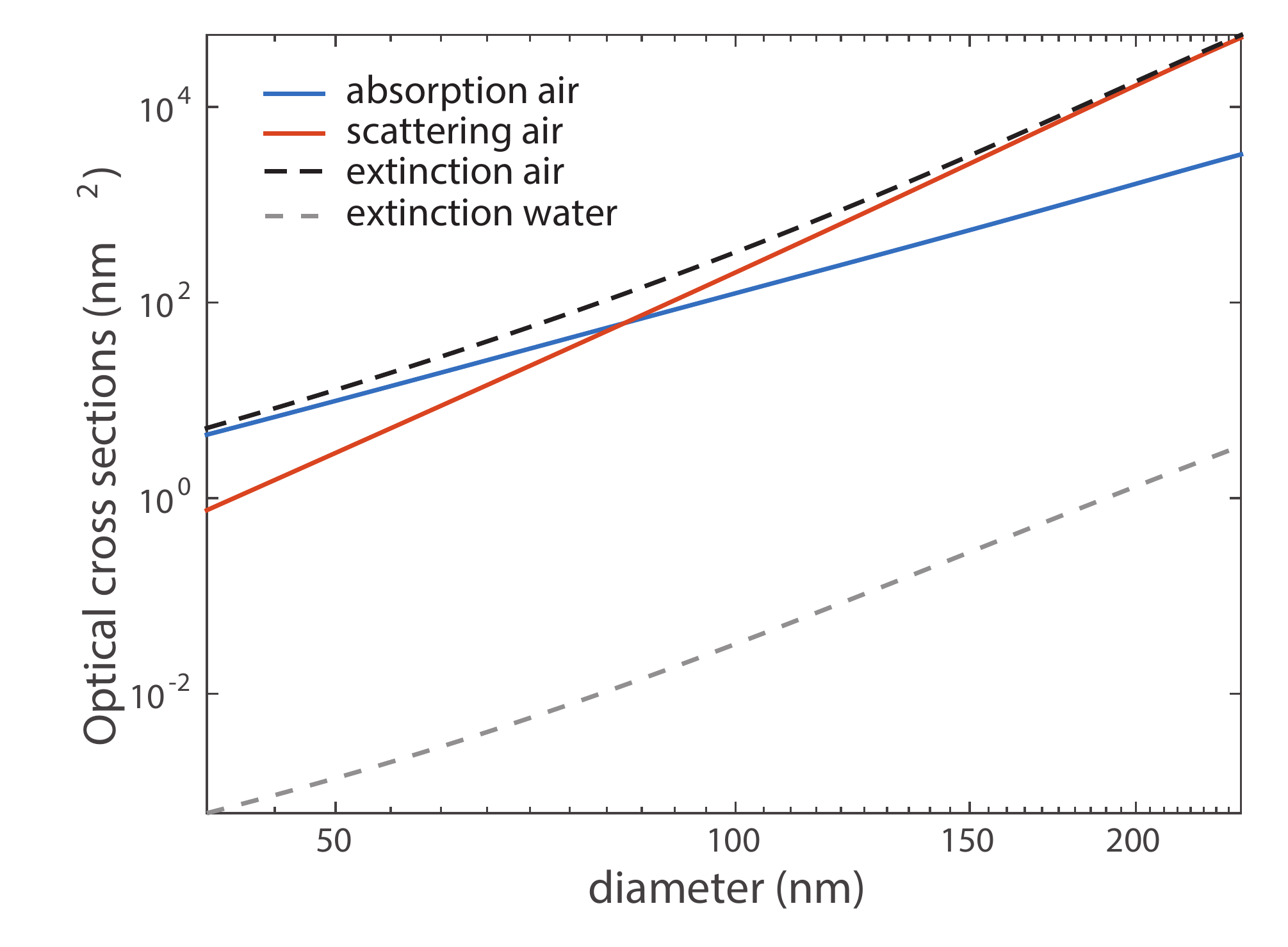}
  \captionof{figure}{Absorption, scattering, and extinction cross sections as a function of particle diameter calculated in air by Mie theory \cite{Mie1908,Bendix2010}, for comparison the extinction cross section for gold in water is also shown.}
\end{center}

\subsection*{Emission by nanoparticles under laser excitation}
For the system described in Ref. \cite{Jauffred2015}, gold nanoparticles are individually trapped with a NIR laser beam (1064 nm) under atmospheric pressure. However, for simplicity, imagine an ensemble of very small gold particles in vacuum irradiated with a laser beam. In this case, the gravity is negligible compared to the interparticle electrostatic interactions in this colloidal system. Its apparent density is very small, which indicates that particles only seldom are in direct contact. Consequently, the thermal conductivity of the system is very low. Thus, when irradiated, the absorbed power causes an increase in temperature and a corresponding heat flux. Hence, the energy balance equation of a single gold nanoparticle can be written as \cite{Roura2002}:
\begin{equation}
P_{abs}(I_L) = P_{em}(T ) + c_p\frac{dT}{dt}
\end{equation}
where $I_L$ is the laser intensity, $P_{abs}$ and $P_{em}$ are the power absorbed from the laser and dissipated by the particles, respectively. $T$ is the temperature of the particles and $c_p$ the heat capacity. The absorbed power will be proportional to the laser intensity, according to 
\begin{equation}
P_{abs} = AI_L,
\end{equation}
where $A$ depends on geometrical factors like the shape and size of the particles and on optical parameters like the absorption and scattering cross sections given in Fig. 1. As the particle are in low numbers, the heat conduction from particle to particle is neglected. Therefore, in vacuum  the only mechanism able to dissipate heat from the particles is blackbody thermal radiation and the emitted power will follow Stefan-Boltzmann law \cite{Roura2002}:
\begin{equation}
P_{em} = B\sigma_B(T^4-T_R^4),
\end{equation}
\begin{center}
  \includegraphics[width=8cm]{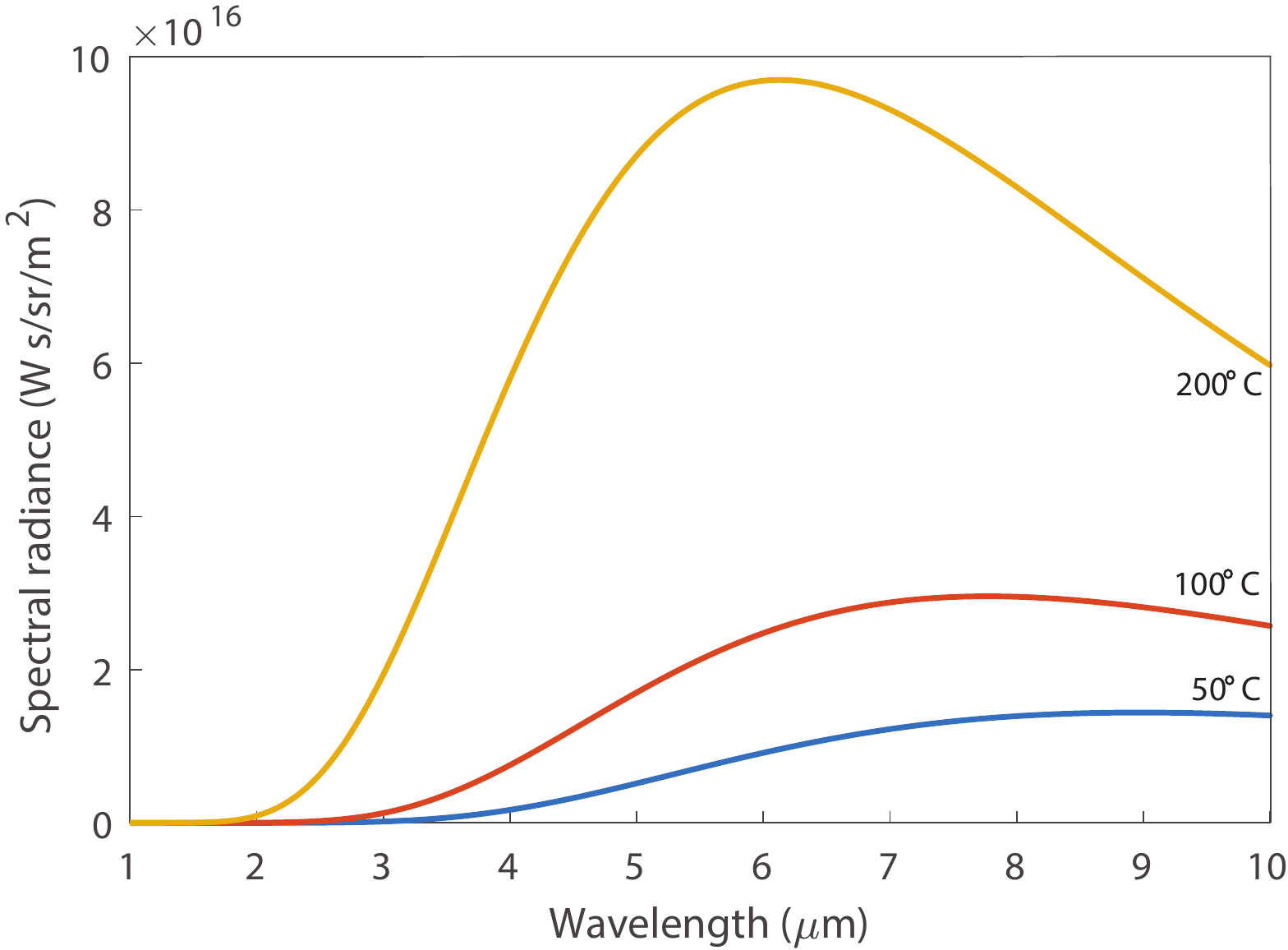}
  \captionof{figure}{Spectral radiance over the NIR spectrum for 50$^{\circ}$ C, 100$^{\circ}$ C, and 200$^{\circ}$ C.}
\end{center}
where $\sigma_B$ is Stefan-Boltzmann's constant, $T_R$ the room temperature and $B$ a constant that depends on the emissivity and geometry of particles. This blackbody emission can be detected by integrating over all emitted wavelengths. Alternatively, the emission in a narrow spectral range around a given wavelength, $\lambda$, can be collected through a monochromator. In this case, the measured intensity will follow Planck's spectral radiance:
\begin{equation}
I_{em}(\lambda)=\epsilon\frac{2\pi h c^2}{\lambda ^5(e^{hc\lambda k_BT}-1)},
\label{eq:Iem}
\end{equation}
where $\epsilon$ is the emissivity, $h$ is Planck's constant, $c$, the speed of light and $k_B$, Boltzmann's constant. For nanoparticles irradiated and heated to a few hundred degrees (Fig. 2), the peak emission is in the NIR spectrum with a tail into the visible regime.  
\begin{center}
  \includegraphics[width=8cm]{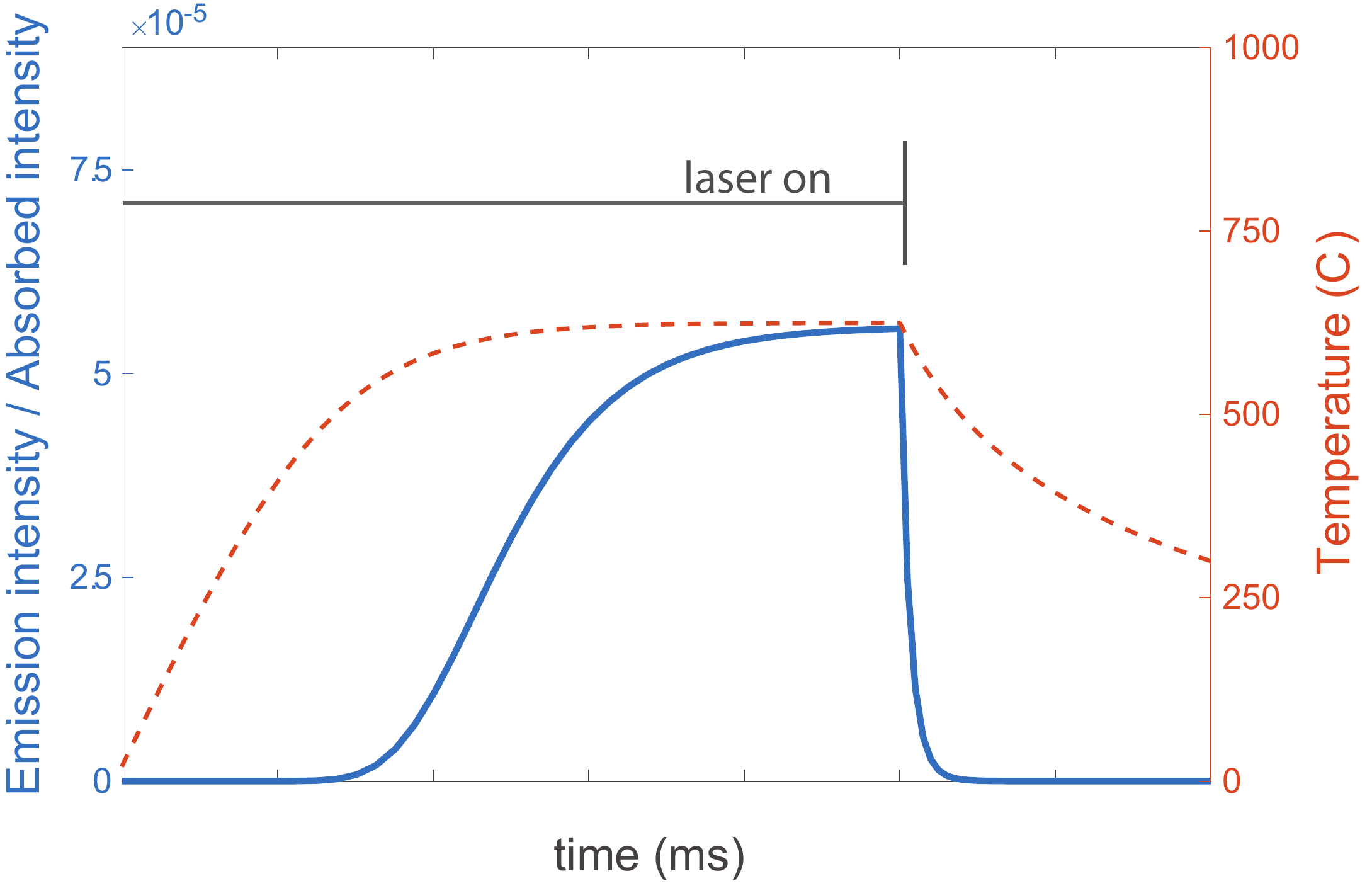}
  \captionof{figure}{Evolution of temperature for gold nanoparticles (200 nm) in vacuum (red dashed curve) irradiated with a 1064 nm NIR laser beam and the corresponding emitted thermal radiation at 470 nm wavelength (blue solid curve)}
\end{center}

Following the train of thought of Ref. \cite{Roura2002} I used this set of equations to calculate the radiation emitted by spherical gold nanoparticles ($R=100$ nm). In order to simplify the problem, I considered every particle to absorb and emit radiation independently, i.e., neglecting shadowing effects. Under these ideal conditions, the constants $A$ and $B$ are simply: 
\begin{equation}
A = \pi R^2(1-e^{-\alpha R}) 
\end{equation}
and 
\begin{equation}
B = 4\pi R^2(1-e^{-\alpha R}),
\end{equation}
where $R$ is the radius of the particles, $\alpha$ the mean optical absorption and the quantity in parentheses corresponds to the emissivity of an infinite layer of thickness $R$. The heating of the particles under irradiation with a NIR 1,064 nm laser was calculated, with parameters detailed in table 1. The temperature evolution has been plotted in Fig. 3 (red curve). Initially, when the laser is turned on, the temperature increases at constant rate, as the emitted power is small. Hence, the heating rate is proportional to the intensity of the laser. This behavior continues up to $\sim500^{\circ}$ C as thermal emission is only important at high temperatures. When the laser is turned off, the aerosol cools proportional to the emitted power, i.e., $T^4-T_R^4$. Once the temperature evolution versus time is known, it is possible to calculate the intensity of radiation emitted at any wavelength by simply introducing $T$ in Planck's distribution Eq. (\ref{eq:Iem}). The emitted intensity at a wavelength of 470 nm (blue) has been calculated and the result is shown in Fig. 3 (solid blue curve). Interestingly, the emitted intensity only raises when temperatures has reached $\sim500^{\circ}$, because of the non-linear dependence on $T$ given by the Planck distribution. In contrast, the emission decays rapidly when the laser is turned off. 
\begin{table}
\caption{Parameters}
\begin{tabular}{l l}\hline
%Laser wavelength & $\lambda = 1064$ nm\\
Laser intensity & $I_L = 146$ mW/mm$^2$\\
Room temperature & $T_R = 293$ K\\ 
Particle radius & $R = 100$ nm\\
Particle mass & $m = 8.0927\cdot 10^{-14}$ g\\
Heat capacity per particle & $c_p = 1.0440\cdot 10^{-14}$ J/K $^{a}$\\
Emissivity & $\varepsilon = (1-\exp(-\alpha R))$\\
Absorption coefficient & $\alpha = 7.8287\cdot 10^{7}$ m$^{-1}$ $^{b}$\\
Reflectivity & $r = 0$\\
\hline
\end{tabular}
\newline
{\footnotesize{$^{a}$ $c_p$=$m\cdot 0.129$ J/gK at room temperature from Ref. \cite{Takahashi1986}.}}\\
{\footnotesize{$^{b}$ for $\lambda$=1064 nm \cite{Rakic1998}.}}
\end{table}

According to this analysis, the laser beam is able to heat the particles because of the lack of dissipation mechanisms. This is why in vacuum; only radiative thermal emission is significant. However, at atmospheric pressure, heat conduction through the surroundings would be so efficient that no emitted radiation at all would be detected. This means that at intermediate pressures one would detect a progressive diminution of the emitted radiation. This can be easily calculated. At steady state, the energy balance equation will be:
\begin{equation}
P_{abs}(I_L) = P_{em}(T) + P_{gas}(T),
\label{eq:Pabs}
\end{equation}
where $P_{gas}(T)$ is the power dissipated through the gas. At a low enough pressure, it is approximately the product of the number of gas collisions on the particle surface
times the mean energy exchanged in one collision \cite{Roura1995}:
\begin{equation}
P_{gas}\approx 4\pi R^2\frac{p}{\sqrt{4\pi mk_BT_R}}k_B(T-T_R)\frac{3}{2},
\label{eq:Pgas}
\end{equation}
where $p$ and $T_R$ are the gas pressure and temperature, respectively, $m$ is the atomic mass and the factor 3/2 arises from the assumption of a monoatomic gas. Eq. (\ref{eq:Pabs}) and Eq. (\ref{eq:Pgas}) allow us to calculate the dependence of the steady state emission intensity versus gas pressure. The result is an exponential dependence:
\begin{equation}
I_{em} = I_0e^{-p/p_0},
\end{equation}
with a very conservative choice of $p_0$ to be 100 Pascals \cite{Roura2002}. With this and the parameters listed in table 1, the emission intensity for atmospheric pressure ($\sim$100 kPa) is less than $10^{-300}$ of the emitted intensity in vacuum (Fig. 4). 
\begin{center}
  \includegraphics[width=8cm]{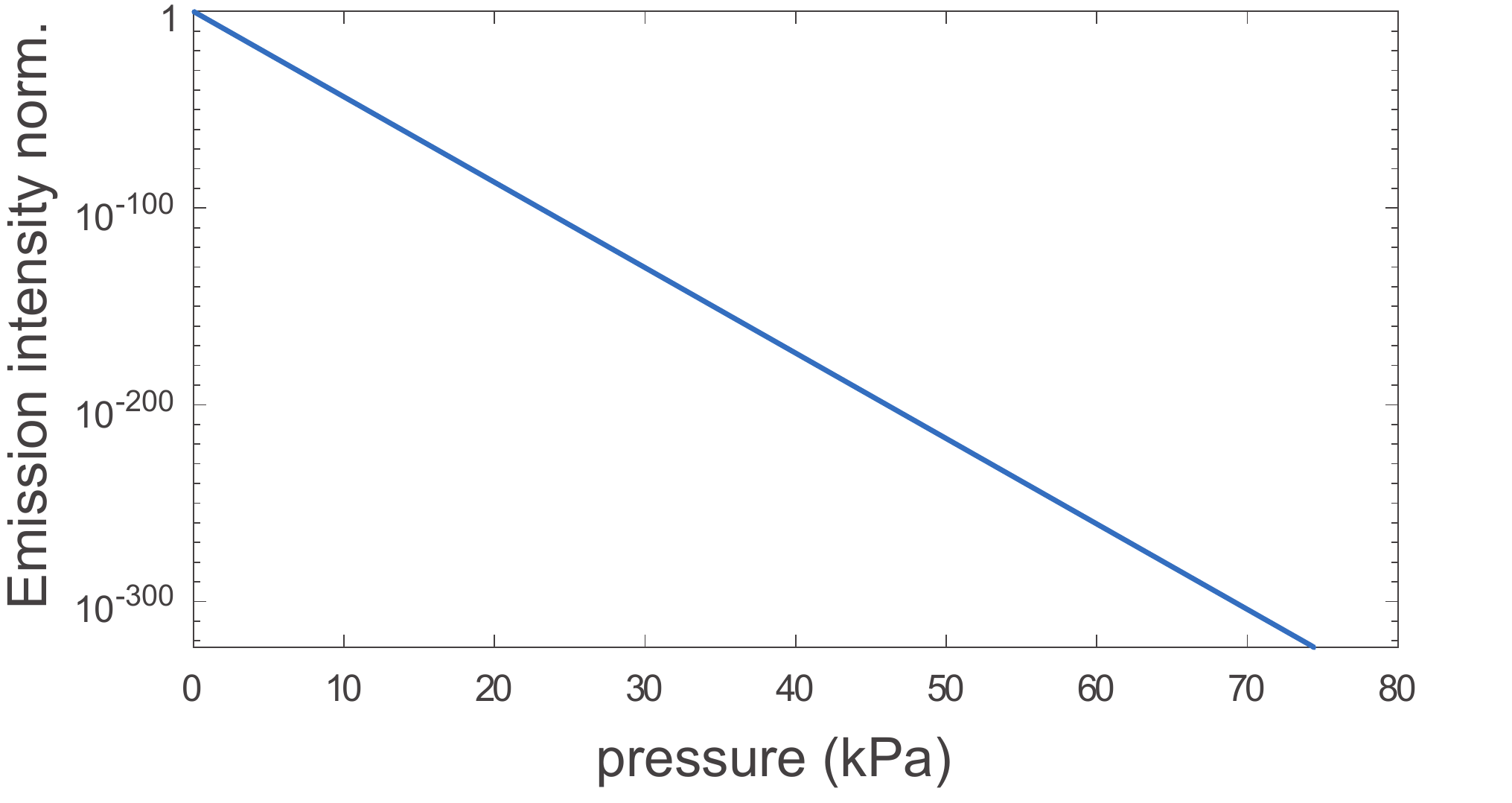}
  \captionof{figure}{The dependence of the steady state emission intensity on the surrounding gas pressure.}
\end{center}

\subsection*{Concluding remarks}
Thermal (blackbody) radiation has a spectrum that depends entirely on the temperature of the particle and the emission intensity for a specific wavelength can be calculated from Planck's law. Therefore, by balancing the absorbed power with the emission power and the heat dissipation, the particle temperature can be extracted. However, at atmospheric pressure most of the absorbed power is dissipated in the surrounding gas. Furthermore, standard thermal imaging, which is often used to measure heating of nanoparticles in suspension \cite{Cole2009,Pattani2013,Jonsson2014} does not apply for nanothermometry. The reason is that the wavelength of several microns of the peak intensity, given by Planck' law, would lead to a very poor spatial resolution. Furthermore, as the emission only becomes pronounced for temperatures of several hundreds of degrees, the measurable temperature range is limited to temperatures far above 100$^{\circ}$ C. Thus, this method is not appropriate to measure ambient temperature changes of single nanoparticles. 

Thermal spectroscopy for nanothermometry is further challenged by the fact that most optical components does not transmit/reflect light with the same probability over a wavelength range that is large enough for spectroscopy, e.g. visible light. For these reasons, thermal spectroscopy for nanothermometry should not be the first option if the goal is to measure the temperature of gold nanoparticle aerosols that are both optically trapped and heated by a single laser. 
%%%%%%%%%%%%%%%%%%%%%%%%%%%%%%%%%%%%%%%%%%%%%%%%%%%%%%%%%%%%%
%\bibliographystyle{achemso}
%\bibliography{../../../Bibtex/ChemRev2017}
\providecommand{\latin}[1]{#1}
\providecommand*\mcitethebibliography{\thebibliography}
\csname @ifundefined\endcsname{endmcitethebibliography}
  {\let\endmcitethebibliography\endthebibliography}{}

\end{document}